\pdfoutput=1
\documentclass[12pt]{article}
\usepackage{amsmath}

\usepackage{amssymb}
\usepackage{graphicx}
\usepackage{adjustbox}
\usepackage{multirow}
\usepackage{booktabs,caption}
\usepackage[flushleft]{threeparttable}
\usepackage{float}
\newcommand{\be}{\begin{equation}}
\newcommand{\ee}{\end{equation}}
\newcommand{\bea}{\begin{eqnarray}}
\newcommand{\eea}{\end{eqnarray}}

\catcode`@=12

\begin{document}
\thispagestyle{empty}
\begin{center}
{\Large\bf
{Gravitational wave signatures from an extended inert doublet dark matter model}}\\
\vspace{1cm}
{{\bf Avik Paul}\footnote{email: avik.paul@saha.ac.in},
{{\bf Biswajit Banerjee}\footnote{email: biswajit.banerjee@saha.ac.in }},
{\bf Debasish Majumdar}\footnote{email: debasish.majumdar@saha.ac.in}}\\
\vspace{0.25cm}
{\normalsize \it Astroparticle Physics and Cosmology Division,}\\
{\normalsize \it Saha Institute of Nuclear Physics, HBNI,} \\
{\normalsize \it 1/AF Bidhannagar, Kolkata 700064, India}\\
\vspace{1cm}
%%%%%%%%%%%%%%%%%%%%%%%%%%%%%%%%%%%%%%%%%%%%%%%%%%
%{\bf ABSTRACT}
%%%%%%%%%%%%%%%%%%%%%%%%%%%%%%%%%%%%%%%%%%%%%%%%%%
\end{center}
\begin{abstract}
%20190720:
We consider a particle dark matter model by extending the scalar
sector of the Standard
Model by an additional SU(2) scalar doublet which is made ``inert"
(and stable)
by imposing a discrete $Z_2$ symmetry under which the additional
scalar doublet is odd (and the SM is even) and it does not develop
any vacuum expectation value (VEV). The lightest inert particle (LIP)
of this inert doublet model (IDM) can be a viable candidate for
Dark Matter. The IDM model is further extended by an additional
singlet scalar which is also even under $Z_2$ and develop a VEV
on spontaneous symmetry breaking. This additional scalar singlet
mixes with SM Higgs and on diagonalisation of the mass matrix
two CP even scalar eigenstates are obtained one of which is attributed
to the physical Higgs (with mass 125.09 GeV). The LIP is the dark matter
candidate in the extended model. For such a particle dark matter model we
explore the first-order electroweak phase transition and consequent
production of Gravitational Waves (GW) at that epoch of the
early Universe and calculate the intensities and frequencies for such waves. We then investigate the detection possibilities of such GWs at the future spaceborne primordial GW detectors such as eLISA, BBO, ALIA, DECIGO, U-DECIGO and ground-based detector aLIGO.
\end{abstract}
%\hspace{1cm}\keywords{Dark matter, Gravitational Wave}
\newpage
\section{Introduction}
Recently, with the detection of a gravitational wave (GW) event \cite{LIGOScientific:2016,{LIGOScientific:2017}} from a black hole binary merger with the Laser Interferometer Gravitational-Wave Observatory (LIGO) the era of GW-astronomy has begun.
The production of GW can also be associated with several other different mechanisms such as topological defects of the domain walls and cosmic strings \cite{Vilenkin:2000}, inflationary quantum fluctuations \cite{Starobinsky:1979ty}, preheating \cite{Khlebnikov:1997di}, first-order phase transitions in the early Universe \cite{Witten:1984rs, {Hogan:1986qda}}, etc.
In this work, we explore production of GW from the first-order phase transitions in the early Universe for a proposed dark matter particle physics model that is constructed by simple extension of  Standard Model (SM) by an inert doublet and a scalar singlet. Also a strong first-order electroweak phase transition helps to explain electroweak baryogenesis \cite{Kuzmin:1985}-\cite{Morrissey:2012}. It is to be noted that 
the first-order cosmological phase transition originates from the bubble nucleation of a true vacuum state at a temperature known as the nucleation temperature at which the probability for a single bubble to nucleate within the horizon volume is of the order one. Initially, the bubbles are considered to have all possible shapes with different surface tension and internal pressure.
The bubbles with the size just large enough for avoiding collapse are considered as a critical bubble. 
The bubbles which are smaller than the critical bubble tend to collapse whereas the larger bubbles tend to expand due to the pressure difference between the false and true vacua. 
During the collisions, the bubbles cannot retain their spherical symmetry which initiates phase transitions and as a result, gravitational waves are produced.
The details of the GW production mechanisms are discussed in \cite{Kosowsky:1992a}-\cite{Caprini:2009}. The GWs are produced from the strong first-order electroweak phase transition mainly by the following three mechanisms: bubble collisions \cite{Kosowsky:1992a}-\cite{Caprini:2008}, sound waves in the hot plasma \cite{Hindmarsh:2014}-\cite{Hindmarsh:2015} and magnetohydrodynamic turbulence of bubbles \cite{Caprini:2006}-\cite{Caprini:2009} in the early Universe.

The electroweak phase transition as explained by SM of particle physics initiated by spontaneous symmetry breaking through Higgs mechanism when the Higgs acquires a vacuum expectation value (VEV) is a smooth crossover and not a first-order phase transition \cite{Kajantie:1996}-\cite{Csikor:1999}. However extension of SM by adding scalar singlet field or a Higgs like doublet can induce strong first-order phase transitions. The phase transitions can be of two types: a) a one-step process, which involves only initial and final phases and b) a two or multi-step processes, which involves one or more intermediate metastable phases along with the initial and final phases \cite{Land:1992}-\cite{Blinov:2015}. In order to explore the gravitational wave production from the first-order phase transition, many authors have considered different types of particle dark matter (DM) models \cite{Kozaczuk:2014kva}-\cite{Huang:2017rzf}. 
%So far, different types of particle candidate of dark matter have been proposed such as Weakly Interacting Massive Particles (WIMPs) \cite{Jungman:1995df}-\cite{Biswas:2013nn}, Febbly Interating Massive Particles (FIMPs) \cite{Yaguna:2011qn,{Molinaro:2014lfa}}, Axions \cite{Peccei:2006as}-\cite{Paul:2018msp}, neutralino \cite{Jungman:1995df}, Fuzzy dark matter \cite{Lidz:2018fqo,{{Amorisco:2018dcn}}}, Kaluza Klein dark matter \cite{Bergstrom:2006ny}, from extra dimensional models \cite{Hsieh:2006qe} etc. 
%In this paper, we consider WIMP as the particle candidate of dark matter. In the early Universe, WIMPs are believed to be produced thermally and were in thermal equilibrium. The WIMPs decoupled from the thermal bath when the temperature of the Universe dropped below the WIMPs mass. This is known as `decoupling' after which the WIMPs are considered as a relic particle as well as a particle candidate of cold-dark matter (CDM). 
In this work, we extend the SM by adding an extra Higgs doublet and a real singlet scalar. The added doublet is an inert doublet \cite{Ma:2006}-\cite{Lopez:2011} in the sense that it does not have any direct coupling with fermion. A $Z_2$ symmetry is imposed to make it stable.
%The IDM can be further extended with an additional real singlet scalar \cite{Banik:2014cfa,{Banik:2014eda}}.
The lightest stable inert particle is attributed to a viable particle candidate of dark matter. The extra singlet scalar mixes with the SM Higgs. This model has already been discussed in previous works \cite{Banik:2014cfa,{Banik:2014eda}}.
We demonstrate that this model, in addition to provide a viable particle candidate, induces  
strong first-order electroweak phase transition. Moreover, introducing a new scalar particle increases the degrees of freedom in the thermal plasma and improves the strength of the electroweak phase transition. In this work, the mass of the second physical scalar $h_{2}$, appearing due to the interaction between singlet scalar and the SM Higgs has been considered heavier as well as lighter than the Higgs mass. We constrain the model parameters by using vacuum stability \cite{Kannike:2012pe}, perturbativity,  Large Electron-Positron Collider (LEP)\cite{Beringer:2012} and Large Hadron Collider (LHC) bounds, PLANCK bound on the DM relic density \cite{Ade:2013zuv} and the limits given by spin independent direct detection experiments like XENON-1T \cite{Aprile:2015uzo}, PandaX-II  \cite{Tan:2016zwf}, LUX \cite{Akerib:2016vxi} and DarkSide-50 \cite{Marini:2016haq}. We also use limits from future generation colliders such as High Luminosity LHC (HL-LHC), International Linear Collider (ILC) and China Electron Positron Collider (CEPC) or TLEP \cite{Profumo:2014opa}. We choose some benchmark points (BPs) from the allowed parameter space to calculate the GW production due to the first-order phase transition induced by the present model. We also discuss the detectability of such GW by the future space interferometers such as Big Bang Observer (BBO) \cite{Moore:2015,{Corbin:2006}}, Evolved Laser Interferometer Space Antenna (eLISA) \cite{Moore:2015}, Advanced Laser Interferometer Antenna (ALIA) \cite{Gong:2015}, DECi-hertz Interferometer Gravitational wave Observatory (DECIGO) \cite{Moore:2015,{Musha:2017}}, Ultimate-DECIGO (U-DECIGO) \cite{Kudoh:2006} and ground-based detector Advanced LIGO (aLIGO) \cite{Moore:2015,{aLIGO:2019}}.

The paper is organised as follows. In Sect. 2, we present the extension of an inert doublet model by introducing a singlet scalar and derive the relations between model parameters. In Sect. 3, we discuss both the theoretical and experimental bounds that we have used to constrain the model parameter space. The calculations of relic density and direct detection cross-section of this extended inert doublet model (IDM) are given in Sect. 4. In this section, we also discuss the viable model parameter space from all the constraints mentioned in Sect. 3. In Sect. 5, we present the finite temperature effective potential to study the electroweak phase transitions in our model. The production mechanisms of GWs from the first-order phase transitions are also furnished in this section. Finally, we summarize and conclude our work in Sect. 6.

\section{The Model}
In this work we extend the SM of particle physics by an extra Higgs doublet $\Phi_I$ and a real singlet scalar $S$.
% Here we further extended the IDM model with an additional real singlet scalar.
While $\Phi_I$ is $Z_2$ odd, the SM and the other added scalar singlet is $Z_2$ even. The extra doublet does not acquire any VEV, while the added singlet acquires a VEV on spontaneous symmetry breaking and mixes with SM Higgs. The dark matter candidate is the lightest of the two neutral scalars of the inert doublet. The potential of the scalar sector of the model can be expressed as

\begin{equation}\label{eq:1}
\begin{aligned}
V=m_1^2 \Phi_H^\dagger\Phi_H + m_2^2 \Phi_I^\dagger\Phi_I+\dfrac{1}{2}m_s^2 S^2 + \lambda_1 \left(\Phi_H^\dagger\Phi_H \right)^2 + \lambda_2 \left(\Phi_I^\dagger\Phi_I \right)^2 + \lambda_3 \left(\Phi_H^\dagger\Phi_H \right) \left(\Phi_I^\dagger\Phi_I \right) \\+ \lambda_4 \left(\Phi_I^\dagger\Phi_H \right) \left(\Phi_H^\dagger\Phi_I \right) + \dfrac{\lambda_5}{2} \left[\left(\Phi_I^\dagger\Phi_H \right)^2+\left(\Phi_H^\dagger\Phi_I \right)^2 \right] + \rho_1 \left(\Phi_H^\dagger\Phi_H \right) S \hspace{2.7cm} \\+\rho_1^{\prime} \left(\Phi_I^\dagger\Phi_I \right) S + \rho_2 \left(\Phi_H^\dagger\Phi_H \right) S^2+ \rho_2^{\prime} \left(\Phi_I^\dagger\Phi_I \right) S^2 + \dfrac{\rho_3}{3} S^3+ \dfrac{\rho_4}{4} S^4. \hspace{3.5cm}
\end{aligned}
\end{equation}
As mentioned, after spontaneous symmetry breaking SM Higgs field $\Phi_H$ acquires a non zero VEV $v=$246.22 GeV and also the scalar $S$ acquires a VEV $v_s$. The SM Higgs, additional Higgs doublet and the scalar particle is represented as
\begin{equation}\label{eq:2}
\phi_H=\left(
\begin{array}{c}
 G^+\\
 \dfrac{1}{\sqrt{2}} \left(v+h+iG_0\right)\\
\end{array}
\right), \hspace{1mm}\phi_I=\left(
\begin{array}{c}
 H^+ \\
 \dfrac{1}{\sqrt{2}} \left(H_0+iA_0 \right)\\
\end{array}
\right), \hspace{1mm}S=v_s+s,
\end{equation}
where $G^+$ and $G_0$ are the charged and neutral Goldstone bosons respectively. After SSB the Goldstone bosons are absorbed in the gauge bosons ($W^{\pm},Z$) \cite{Banik:2014eda}.
After minimising the scalar potential represented in Eq. (\ref{eq:1}) using the conditions
\begin{equation}\label{eq:3}
\dfrac{\partial V}{\partial h}\Big|_{h=s=H^+=H_0=A_0=0}=\dfrac{\partial V}{\partial s}\Big|_{h=s=H^+=H_0=A_0=0}=0,
\end{equation}
we obtain
\begin{equation}\label{eq:4}
\begin{aligned}
m_1^2+\lambda_1v^2+\rho_1v_s+\rho_2v_s^2=0,\\
m_s^2+\rho_3v_s+\rho_4v_s^2+\dfrac{\rho_1v^2}{2v_s}+\rho_2v^2=0.
\end{aligned}
\end{equation}
The mass matrix of the scalar sector is obtained by calculating the second-order derivatives $\left(\dfrac{\partial ^2V}{\partial {h}^2},\hspace{1mm} \dfrac{\partial ^2V}{\partial {s}^2}, \hspace{1mm}\dfrac{\partial ^2V}{\partial {H^{\pm}}^2},\hspace{1mm} \dfrac{\partial ^2V}{\partial {H_0}^2},\hspace{1mm} \dfrac{\partial ^2V}{\partial {A_0}^2},\hspace{1mm} \dfrac{\partial ^2V}{\partial {h}\partial {s}}\right)$ of the scalar potential (Eq. (\ref{eq:1})) and the elements are

\begin{equation}\label{eq:5}
\mu_{h}^2=m_1^2+3\lambda_1 v^2+\rho_1 v_s+\rho_2 v_s^2=2\lambda_1 v^2,
\end{equation}

\begin{equation}\label{eq:6}
\mu_s^2=m_s^2+\rho_2 v^2+2\rho_3 v_s+3\rho_4 v_s^2=\rho_3v_s+2\rho_4v_s^2-\dfrac{\rho_1v^2}{2v_s},
\end{equation}

\begin{equation}\label{eq:8}
m_{H^\pm}^2=m_2^2+\dfrac{\lambda_3v^2}{2}+\rho_1^\prime v_s+\rho_2^\prime v_s^2,
\end{equation}

\begin{equation}\label{eq:9}
m_{H_0}^2=m_2^2+\left(\lambda_3+\lambda_4+\lambda_5\right)\dfrac{v^2}{2}+\rho_1^\prime v_s+\rho_2^\prime v_s^2,
\end{equation}

\begin{equation}\label{eq:10}
m_{A_0}^2=m_2^2+\left(\lambda_3+\lambda_4-\lambda_5\right)\dfrac{v^2}{2}+\rho_1^\prime v_s+\rho_2^\prime v_s^2,
\end{equation}

\begin{equation}\label{eq:7}
\mu_{hs}^2=\rho_1v+2\rho_2v_sv.
\end{equation}
As $h$ and $s$ mix, we diagonalise the mass matrix in $h$, $s$ basis by a unitary matrix $U$ 
\begin{equation}\label{eq:11}
\left(
\begin{array}{c}
 h_1 \\
 h_2  \\
\end{array}
\right)=U\left(
\begin{array}{c}
 h \\
 s  \\
\end{array}
\right)
=
\left(
\begin{array}{cc}
 \cos\theta & -\sin\theta\\
 \sin\theta & \cos\theta\\
\end{array}
\right)\left(
\begin{array}{c}
 h \\
 s \\
\end{array}
\right),
\end{equation}
to obtain two physical mass eigenstates $h_1$ and $h_2$ as
\begin{equation}\label{eq:12}
h_1=h\hspace{1mm}\cos\theta-s\hspace{1mm}\sin\theta, \hspace{1mm}h_2 =h\hspace{1mm}\sin\theta +s\hspace{1mm}\cos\theta,
\end{equation}
where $\theta$ is the mixing angle that can be computed from
\begin{equation}\label{eq:13}
\tan\theta=\dfrac{y}{1+\sqrt{1+y^2}}, \hspace{2mm} \text{where} \hspace{1mm} y=\dfrac{2 \mu_{hs}^2}{\mu_{h}^2-\mu_s^2} \hspace{1mm}.
\end{equation}
The expressions for the mass eigenstate of the two physical scalars $h_1$ and $h_2$ are given as
\begin{equation}\label{eq:14}
m_{h_1, h_2}^2=\dfrac{\left(\mu_{h}^2+\mu_s^2\right)}{2}\pm \dfrac{\left(\mu_{h}^2-\mu_s^2\right)}{2} \sqrt{1+y^2},
\end{equation}
where the `+' sign is for $h_1$ and `$-$' sign is for $h_2$. In the present work, $h_1$ is attributed to the SM like Higgs boson with mass $m_{h_1}=125.09$ GeV \cite{Patrignani:2016xqp} and $h_2$ is the other scalar with mass $m_{h_2}$.  Here we consider both the cases when $m_{h_2}>m_{h_1}$ and when $m_{h_2}<m_{h_1}$. Considering the coupling $\lambda_5$ (in Eq. (\ref{eq:1})) to be less than zero, we get $H_0$ to be the lightest stable particle and the dark matter candidate in our present work. Using Eqs. (\ref{eq:4})-(\ref{eq:14}) we obtain the following relations
\begin{equation}\label{eq:15}
\lambda_1=\dfrac{m_{h_2}^2\sin ^2\theta+m_{h_1}^2 \cos ^2\theta}{2 v^2},
\end{equation}

\begin{equation}\label{eq:16}
\rho_4=\dfrac{1}{2v_{s}^2}\left(m_{h_2}^2\cos ^2\theta+m_{h_1}^2 \sin ^2\theta + \dfrac{v^2\rho_1}{2v_s}-v_s\rho_3\right),
\end{equation}

\begin{equation}\label{eq:17}
\rho_2=\dfrac{m_{h_1}^2-m_{h_2}^2}{4 v\hspace{.1mm} v_{s}}\sin {2\theta}-\dfrac{\rho_1}{2v_s},
\end{equation}

\begin{equation}\label{eq:18}
m_1^2=-\lambda_1v^2-\rho_1v_s-\rho_2v_s^2,
\end{equation}

\begin{equation}\label{eq:19}
m_2^2=m_{H_0}^2-\dfrac{v^2}{2}\left(\lambda_3+\lambda_4+\lambda_5\right)-\rho_1^\prime v_s -\rho_2^\prime v_s^2,
\end{equation}

\begin{equation}\label{eq:20}
m_s^2=-\rho_3v_s-\rho_4v_s^2-\dfrac{\rho_1 v^2}{2v_s}-\rho_2 v^2.
\end{equation}

\section{Constraints}
In this section various theoretical and experimental bounds are given. These are used to constrain the model parameter space.
\subsection{Theoretical Constraints}
\noindent \underline{\it Vacuum Stability}

From the vacuum stability conditions the bounds on the couplings are given as \cite{Kannike:2012pe}
\begin{equation}\label{eq:21}
\lambda_1, \lambda_2, \rho_4>0, \hspace{1mm} \lambda_3+2\sqrt{\lambda_1 \lambda_2}>0,
\end{equation}

\begin{equation}\label{eq:22}
\lambda_3+\lambda_4-|\lambda_5|+2\sqrt{\lambda_1 \lambda_2}>0, \hspace{1mm}\rho_2+\sqrt{\lambda_1 \rho_4}>0,
\end{equation}

\begin{equation}\label{eq:23}
\rho_2^\prime+\sqrt{\lambda_2 \rho_4}>0,
\end{equation}

\begin{equation}\label{eq:24}
2\rho_2\sqrt{\lambda_2}+2\rho_2^\prime\sqrt{\lambda_1}+\lambda_3\sqrt{\rho_4}+2\left(\sqrt{\lambda_1 \lambda_2 \rho_4}+\sqrt{\left(\lambda_3+2\sqrt{\lambda_1 \lambda_2}\right)\left(\rho_2+\sqrt{\lambda_1 \rho_4}\right)\left( \rho_2^\prime+\sqrt{\lambda_2 \rho_4}\right)}\right)>0,
\end{equation}

\begin{equation}\label{eq:25}
\begin{aligned}
2\rho_2\sqrt{\lambda_2}+2\rho_2^\prime\sqrt{\lambda_1}+\left(\lambda_3+\lambda_4-\lambda_5\right)\sqrt{\rho_4}+2\Big(\sqrt{\lambda_1 \lambda_2 \rho_4} \hspace{3cm}\\+\sqrt{\left(\left(\lambda_3+\lambda_4-\lambda_5\right)+2\sqrt{\lambda_1 \lambda_2}\right)\left(\rho_2+\sqrt{\lambda_1 \rho_4}\right)\left( \rho_2^\prime+\sqrt{\lambda_2 \rho_4}\right)}\Big)>0. \hspace{1cm}
\end{aligned}
\end{equation}

\noindent \underline{\it Perturbativity}

All the quartic couplings in the tree-level potential (Eq. (\ref{eq:1})) must be less than $4\pi$ to be consistent with the perturbative conditions.
\subsection{Experimental Constraints}
\noindent \underline{\it Collider Constraints}

From the LEP experiment, the bound on the model parameter space is given as  \cite{Beringer:2012}
\begin{equation}\label{eq:26}
m_{H_0}+m_{A_0}>m_Z, \hspace{1mm} m_{H^\pm}>79.3 \hspace{1mm} {\text{GeV}}.
\end{equation}

%In the present scenario, we consider $h_1$ is the SM like Higgs boson of mass 125.09 GeV. 
The bounds are also obtained from the LHC experimental results. 
The signal strength of the SM like Higgs $h_1$ in the present model can be expressed as
\begin{equation}\label{eq:27}
R_1=\cos ^4\theta\dfrac{\Gamma ^{\text{SM}}}{\Gamma },
\end{equation}
where $\Gamma ^{\text{SM}}$ and $\Gamma$ are the total SM Higgs decay width and total decay width of SM like Higgs boson of mass 125.09 GeV. The expression of $\Gamma$ can be written as
\begin{equation}\label{eq:28}
\Gamma =\cos ^2\theta \hspace{1mm} \Gamma ^{\text{SM}}+\Gamma ^{\text{inv}},
\end{equation}
where $\Gamma ^{\text{inv}}$ is the invisible Higgs decay width. In our case, there are two possible invisible decay channels of $h_1$, one of them is $\Gamma^{inv}\left(h_1\rightarrow H_0 H_0\right)$ (for  $m_{H_0}\leq m_{h_1}/2$, $m_{H_0}$ being the mass of the dark matter particle $H_0$) and the other one is $\Gamma^{inv}\left(h_1\rightarrow h_2 h_2\right)$ (for $m_{h_2}\leq m_{h_1}/2$) and they are expressed as
 \begin{equation}\label{eq:29}
\Gamma^{inv}\left(h_1\rightarrow H_0 H_0\right)=\dfrac{\left(g_{h_1 H_0 H_0}\right)^2}{16\pi m_{h_1}}\left(1-\dfrac{4m_{H_0}^2}{m_{h_1}^2}\right)^{1/2},
\end{equation}
and
\begin{equation}\label{eq:30}
\Gamma^{inv}\left(h_1\rightarrow h_2 h_2\right)=\dfrac{\left(g_{h_1 h_2 h_2}\right)^2}{16\pi m_{h_1}}\left(1-\dfrac{4m_{h_2}^2}{m_{h_1}^2}\right)^{1/2},
\end{equation}
where the expressions for the couplings $g_{h_1 H_0 H_0}$ and $g_{h_1 h_2 h_2}$ are given as
\begin{equation}\label{eq:69}
g_{h_1 H_0 H_0} = \left( \dfrac{\lambda_L}{2} \cos\theta - \dfrac{\lambda_s}{2} \sin\theta \right)v=\lambda_{h_1 H_0 H_0} v,
\end{equation}
with
\begin{equation}\label{eq:40}
\lambda_L=\lambda_3+\lambda_4+\lambda_5 \hspace{1mm}, \hspace{1mm} \lambda_s=\dfrac{\rho_1^\prime+2\rho_2^\prime v_s}{v},
\end{equation}
and
\begin{equation}\label{eq:71}
\begin{aligned}
g_{h_1 h_2 h_2} = \dfrac{1}{2}(6\lambda_1 v\cos\theta \sin^2\theta + 2\rho_1 \cos^2\theta \sin\theta -\rho_1 \sin^3\theta + 2v\rho_2 \cos^3\theta -4v\rho_2 \cos\theta\sin^2\theta \\+4v_s\rho_2 \cos^2\theta\sin\theta - 2v_s \rho_2 \sin^3\theta - 2\rho_3 \cos^2\theta\sin\theta -6v_s \rho_4 \cos^2\theta\sin\theta). \hspace{1cm}
\end{aligned}
\end{equation}

The invisible decay branching fraction for SM like scalar can be expressed as
\begin{equation}\label{eq:31}
\text{Br}_{\text{inv}}=\dfrac{\Gamma^{\text{inv}}}{\Gamma}.
\end{equation}
We adopt the bounds on the invisible decay branching fraction for SM scalar to be $\text{Br}_{\text{inv}}\leq24\%$ \cite{CERN:2016} (for $m_{h_1}\ge m_{H_0}/2$). 
The limit on the  scalar mixing $\sin\theta\le0.4$ \cite{Robens:2015}-\cite{Dupuis:2016} and the signal strength of SM Higgs $R_1\ge0.84$ \cite{Khachatryan:2015, {Aad:2016}}
is obtained from LHC experimental results. However, study of future collider experiments can
further constrain  the  scalar mixing angle \cite{Profumo:2014opa}. From the study of Ref.~\cite{Profumo:2014opa} 
it is found that High Luminosity LHC (HL-LHC) restricts the mixing angle to be $\cos\theta\geq 
0.94$. Moreover bounds from International Linear Collider (ILC) is found to be 
more stringent. Projected limit on mixing angle is found to be $\cos\theta\simeq 0.98$ 
for ILC-1 operating at $\sqrt{s}=250$ GeV and $\cos\theta\simeq 0.99$ for ILC-3 
at $\sqrt{s}=1$ TeV. The mixing angle is further constrained to $\cos\theta\sim0.997$ 
when China Electron Positron Collider (CEPC) or TLEP sensitivity is taken into account. 
Therefore, we consider $\sin\theta\leq  0.063$ ($\cos\theta\geq 0.998$), consistent with the limits from future 
collider experiments.

\noindent \underline{\it PLANCK constraint on relic density}

The relic density of dark matter candidate $H_0$ must satisfy the PLANCK observational limit for this to be a viable candidate of dark matter. The relic density limit given by PLANCK observation is
\begin{equation}\label{eq:32}
0.1172\leq \Omega _{\text{DM}}{\rm h}^2\leq 0.1226.
\end{equation}
where $\Omega _{\text{DM}}$ is the DM relic density normalised by the critical density of the Universe and ${\rm{h}}$ is the Hubble parameter normalised by a value of 100 Km s$^{-1}$ Mpc$^{-1}$ \cite{Ade:2013zuv}.

\noindent \underline{\it Direct Searches of Dark matter}

Direct detection experiments of dark matter put an upper bound on dark matter nucleon elastic scattering cross-sections for different dark matter masses. In the present work, we consider the results of the following direct detection experiments of dark matter to constrain the model parameter space XENON-1T 
\cite{Aprile:2015uzo}, PandaX-II  \cite{Tan:2016zwf}, LUX \cite{Akerib:2016vxi} and DarkSide-50 \cite{Marini:2016haq}. 

\section{Dark matter phenomenology}
In this section we furnish the dark matter relic density and direct detection scattering cross-section formulas of the present extended inert doublet model with an additional real singlet scalar. These will be used to compute and constrain the model parameter space. 
\subsection{Relic Density}
%\begin{figure}
%\centering
%\includegraphics[width=8cm,height=6cm]{dm3.png}
%\caption{Feynman diagrams for the dominant annihilation channels of the dark matter candidate $H_{0}$.}
%\label{fig:1}
%\end{figure}
In order to calculate the dark matter relic density one needs to solve the Boltzmann equation which can be expressed as  \cite{Kolb:1990vq}
\begin{equation}\label{eq:33}
\frac{dn_{H_0 }}{dt}+3Hn_{H_0}=-\langle\sigma v\rangle\left[n_{H_0 }^2-\left(n_{H_0 }^{\text{eq}}\right){}^2\right],
\end{equation}
where $n_{H_0}$ is the dark matter number density and $n_{H_0}^{\text{eq}}$ is the dark matter number density at thermal equilibrium. In Eq. (\ref{eq:33}), $\langle\sigma v\rangle$ is the thermal average total annihilation cross-section $\sigma$ times the relative velocity $v$ of the dark matter candidate ($H_0$) and $H$ is the Hubble parameter. The expressions for $\langle\sigma v\rangle$ at a temperature $T$ can be written as 
\begin{equation}\label{eq:34}
\left<\sigma v\right>=\dfrac{1}{8m_{H_0}^4 TK_2^2\left(\frac{m_{H_0 }}{T}\right)}\int _{4 m_{H_0 }^2}^{\infty }ds\left(s-4 m_{H_0 }^2\right)\sqrt{s}K_1\left(\frac{\sqrt{s}}{T}\right) \sigma (s),
\end{equation}
where $\sigma (s)$ is the total annihilation cross-section of the dark matter particle $H_0$, $\sqrt{s}$ is the centre of mass energy, $K_1$ and $K_2$ are the first and second-order modified Bessel functions respectively. 
The relic density of the dark matter candidate $H_0$ can be computed using the expressions which are given as
\begin{equation}\label{eq:35}
\Omega_{\rm DM} {\rm h}^2 =2.755\times 10^8 
\left(\dfrac{m_{H_0}}{\rm{GeV}}\right) Y_0,
\end{equation}  
with
\begin{equation}\label{eq:36}
\dfrac{1}{Y_0}=\dfrac{1}{Y_F}+\left(\dfrac{45 G}{\pi}\right)^{-\frac{1}{2}} \int_{T_0}^{T_F} g_{*}^{1/2} \langle\sigma {\rm v}\rangle dT,
\end{equation}
where $Y_F$ is the value of $Y$ at the freeze-out temperature $T_F$, $G$ the universal 
gravitational constant, $g_{*}$ is the degrees of freedom and $T_0$ is the temperature at the present epoch.

For the purpose of computations we used the publicly available packages namely FeynRules \cite{Alloul:2013bka} where the extended IDM is implemented and finally micrOMEGAs \cite{Belanger:2013oya} (which includes the package CalcHEP within its framework) to calculate the relic density.  In order to calculate the total annihilation cross-sections of the dark matter candidate, the annihilation channels
$H_0 H_0\rightarrow W^+ W^-,\hspace{1mm} ZZ,\hspace{1mm} f\bar{f}$ (where $W$, $Z$ denote the gauge bosons and $f$ denotes the SM fermions) %into final state SM particles (quarks, leptons, gauge bosons, Higgs boson) 
as well as $H_0 H_0\rightarrow h_1 h_1,\hspace{1mm}h_2 h_2,\hspace{1mm} h_1 h_2$ via s, t and u channels as also the relevant four-point interactions have been considered. The micrOMEGAs \cite{Belanger:2013oya} has been used to estimate the relic-density for dark matter candidate $H_0$ considering all the annihilation channels mentioned above.
%The Feynman diagrams for the above-mentioned annihilation channels are shown in Figure \ref{fig:1} and the corresponding expressions for annihilation cross-sections are given in the appendix.

\subsection{Direct detection}
The dark matter particle $H_0$ interacts with the SM particles via Higgs exchange. The expression for the spin independent elastic scattering cross-section is given as 
\begin{equation}\label{eq:37}
\sigma_{\text{SI}}=\dfrac{m_N^4}{\pi \left( m_{H_0}+m_N\right)^2} f^2 \left(\dfrac{\lambda_{h_1 H_0 H_0}\cos\theta}{m_{h_1}^2}+\dfrac{\lambda_{h_2 H_0 H_0}\sin\theta}{m_{h_2}^2}\right)^2,
\end{equation}
where $m_N$ is the nucleon mass and $f$ is the nucleon form factor which has been approximated as 0.3 \cite{cline:2015}. In Eq. (\ref{eq:37}) the couplings $\lambda_{h_1 H_0 H_0}$ and $\lambda_{h_2 H_0 H_0}$ can be written as

\begin{equation}\label{eq:38}
\lambda_{h_1 H_0 H_0} = \left( \dfrac{\lambda_L}{2} \cos\theta - \dfrac{\lambda_s}{2} \sin\theta \right),
\end{equation}

\begin{equation}\label{eq:39}
\lambda_{h_2 H_0 H_0} = \left( \dfrac{\lambda_L}{2} \sin\theta + \dfrac{\lambda_s}{2} \cos\theta \right).
\end{equation}

We calculate the dark matter scattering cross-sections using Eqs. (\ref{eq:37})-(\ref{eq:39}) and then check whether they satisfy the direct detection cross-section bounds obtained from DM direct detection experiments such as XENON-1T 
\cite{Aprile:2015uzo}, PandaX-II  \cite{Tan:2016zwf}, LUX \cite{Akerib:2016vxi} and DarkSide-50 \cite{Marini:2016haq}. 
\subsection{Constraining the parameter space}
The parameter space is first constrained with the theoretical bounds given in Sect. 3.1. It is then further constrained by the collider bounds given in Sect. 3.2. The constrained parameter space thus obtained is then used to calculate the relic densities of the dark matter (Sect. 4.1) in this model. The parameter space is finally constrained by comparing these calculated relic densities with PLANCK results (Eq. (\ref{eq:32})).
%the range of the dark matter relic density given by the PLANCK satellite-borne experiment (Eq. (\ref{eq:32})).
We have checked that the scattering cross-sections and their variations with the dark matter masses calculated using this finally constrained parameter space lie below the upper bounds of such variations given by different dark matter direct detection experiments such as  XENON-1T 
\cite{Aprile:2015uzo}, PandaX-II  \cite{Tan:2016zwf}, LUX \cite{Akerib:2016vxi} and DarkSide-50 \cite{Marini:2016haq}.    

\begin{figure}[H]
\centering
\includegraphics[width=7cm,height=5.3cm]{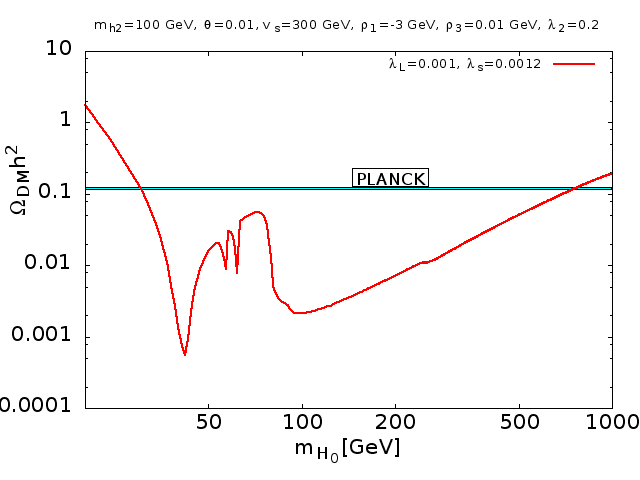}
\includegraphics[width=7cm,height=5.3cm]{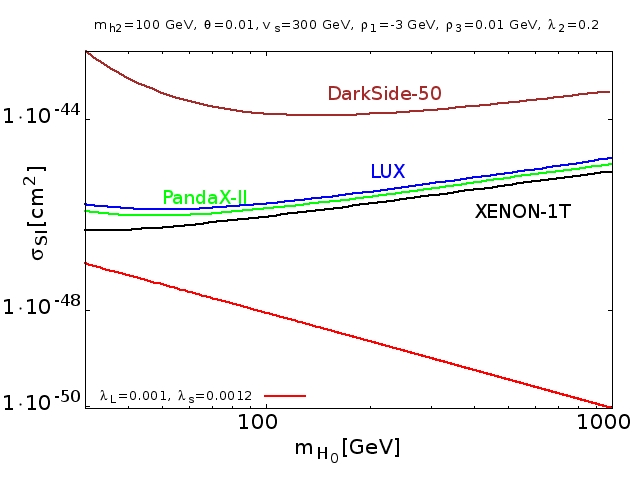}
 \caption{
Variation of DM relic density (left panel) and variation of DM direct detection scattering cross-section (right panel) as a function of DM mass $m_{H_0}$ for BP1. The shadowed area in the left panel bounded by two horizontal parallel lines shows the PLANCK bounds and the red line represents the calculated relic densities. DM direct detection scattering cross-section (right panel), calculated for different DM masses $m_{H_0}$, is overplotted with the bounds given by dark matter direct detection experiments such as XENON-1T, LUX, PandaX-II and DarkSide-50.}
 % Variation of DM relic density as a function of DM mass $m_{H_0}$ (left panel) and Variation of DM direct detection scattering cross-section as a function of DM mass $m_{H_0}$ (right panel) for BP1. The shadowed area bounded by two horizontal parallel lines in the left panel of the Figure shows the PLANCK bounds where the calculated relic density is shown by the red lines. The parameter sets for BP1 are tabulated in Table \ref{t1}.}
 %The left panel is plotted for the variation of relic density of DM as a function of DM mass $m_{H_0}$ and the right panel is for the variation of direct detection scattering cross-section as a function of DM mass $m_{H_0}$. The parameters are used to plot the graphs are tabulated in Table 1 and refereed to as BP1. }
\label{fig:2}
\end{figure}

\begin{figure}[H]
\centering
\includegraphics[width=8cm,height=7cm]{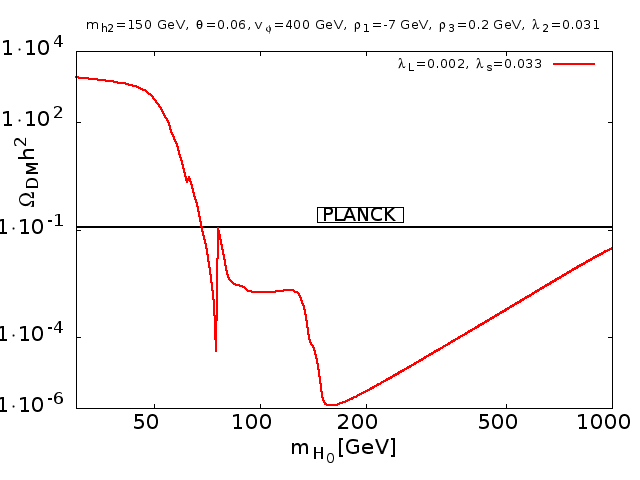}
\includegraphics[width=8cm,height=7cm]{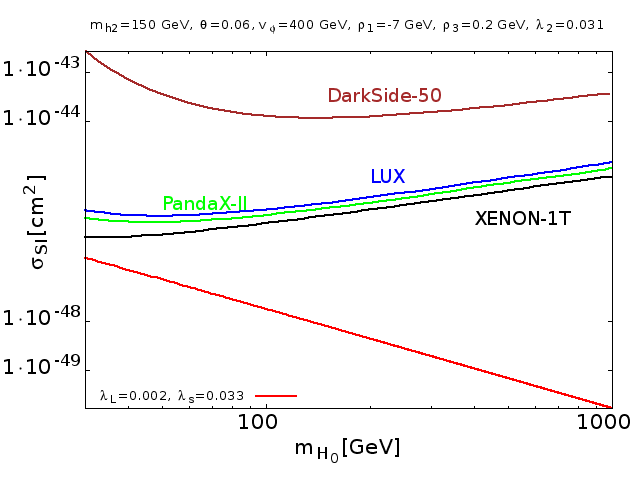}
 \caption{Same as in Figure \ref{fig:2} but for BP2.}
\label{fig:3}
\end{figure}

\begin{figure}[H]
\centering
\includegraphics[width=8cm,height=7cm]{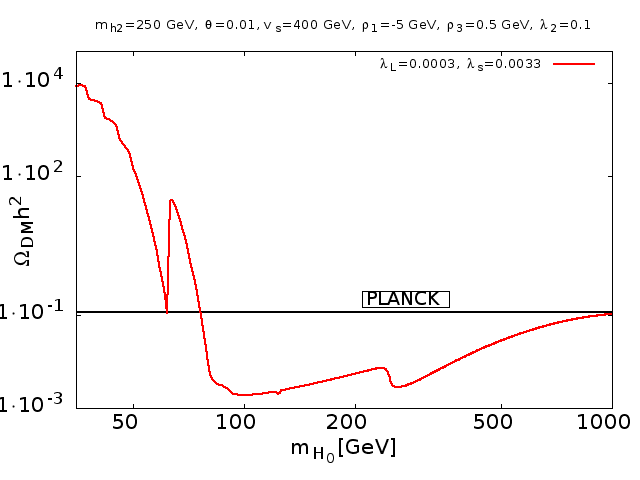}
\includegraphics[width=8cm,height=7cm]{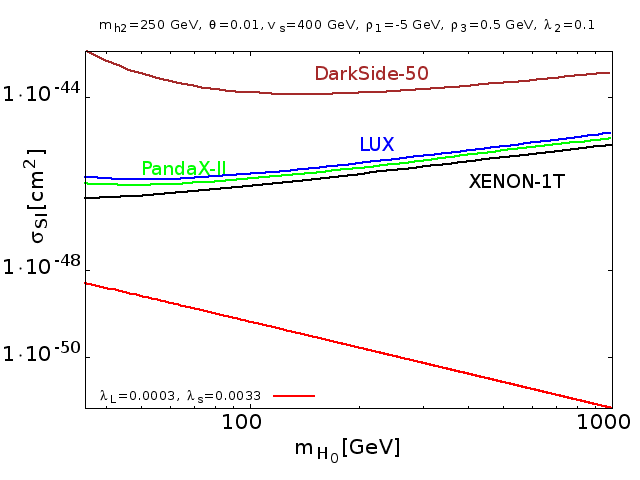}
 \caption{Same as in Figure \ref{fig:2} but for BP3.}
\label{fig:3new}
\end{figure}

\begin{figure}[H]
\centering
\includegraphics[width=8cm,height=7cm]{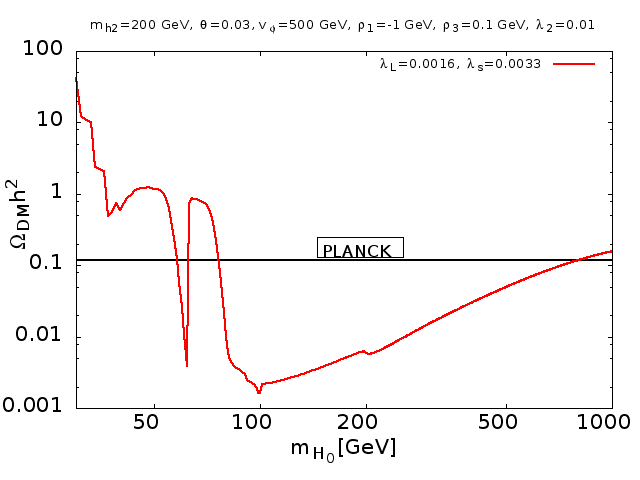}
\includegraphics[width=8cm,height=7cm]{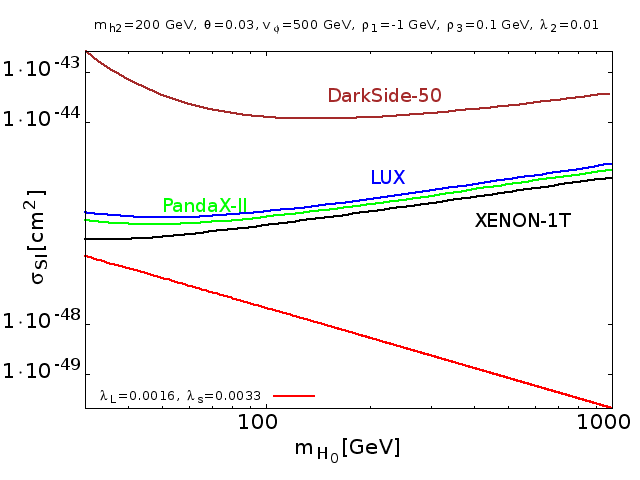}
 \caption{Same as in Figure \ref{fig:2} but for BP4.}
\label{fig:4}
\end{figure}

%For the present work in relation to the generation of gravitational waves from first-order electroweak phase transition with this inert doublet and singlet scalar extended SM, we have chosen four benchmark points (BPs).
In the present work we choose four benchmark points (BPs) from the constrained parameter space and use them to calculate the intensities and frequencies of the gravitational wave that would be produced by this first-order phase transition initiated by the present dark matter model.
These are given in Table \ref{t1}. For the purpose of demonstration, we calculate the relic densities for each of the BPs and show their variations with the mass of the DM candidate $m_{H_0}$ in this model. They are given in left panels of Figures \ref{fig:2}-\ref{fig:4}. As mentioned, the relic densities are computed using micrOMEGAs code \cite{Belanger:2013oya}. We also show in the same Figures (right panels) the variations of dark matter nucleon scattering cross-sections as a function of dark matter masses with the same BPs. 
\begin{table}[H]
\centering
\footnotesize
\begin{tabular}{|l|c|c|c|c|c|c|c|c|c|c|c|r|}
\hline
BP&$m_{H_{0}}$&$m_{h_{2}}$&$v_{s}$&$\sin\theta$&$\rho_{1}$&$\rho_{3}$&$\lambda_{L}$&$\lambda_{s}$&$\lambda_{2}$&$\Omega_{\text{DM}}{\rm h}^2$&$\sigma_{{\rm{SI}}}$\\
 &in GeV&in GeV&in GeV&&in GeV&in GeV&&&&&$\rm{cm^2}$\\
 \hline
1&30&100&300&0.01&-3&0.01&0.001&0.0012&0.2&0.1220&9.41$\times 10^{-48}$\\ 
 \hline
2&68&150&400&0.06&-7&0.2&0.002&0.033&0.031&0.1208&3.69$\times 10^{-48}$\\ 
 \hline
3&62&250&400&0.01&-5&0.5&0.0003&0.0033&0.1&0.1206&1.70$\times 10^{-49}$\\
\hline
4&76&200&500&0.03&-1&0.1&0.0016&0.0033&0.01&0.1195&3.56$\times 10^{-48}$\\
\hline
\end{tabular}
\caption{The chosen four benchmarks points (BPs, BP1-4) to explore the GW production from an extended IDM with an additional real singlet scalar. The relic density and scattering cross-section values for each of the BPs are also mentioned in this Table.}\label{t1}
\end{table}
%For a fixed value of DM mass, from the left panel of Figures \ref{fig:2}-\ref{fig:4}, one can see certain dips in the relic density. This is due to the fact that when a new annihilation channel opens up the annihilation cross-section increases which in turn reduces the relic density. 
The dips in the relic density plots (left panels of Figures \ref{fig:2}-\ref{fig:4}) attribute to the fact that at these dark matter masses the annihilation cross-section suffers a sharpe increase resulting in reduction in relic densities.
From the left panels of Figures \ref{fig:2}-\ref{fig:4}, it is evident that when the DM mass ($m_{H_0}$) is close to ($m_{h_1}/2$) and $m_{h_{2}}/2$, sudden reductions in the relic density occur. This is because at these DM masses annihilation cross-sections increase significantly. For example, for $m_{h_2}=100$ GeV the dip in the relic density for the dark matter mass around 62.5 GeV and 50 GeV are due to the resonant increase of annihilation cross-section. In addition, dips in the relic density plot for DM masses close to $m_W$ (mass of $W$ boson), $m_Z$ (mass of $Z$ boson), $m_{h_1}$ (mass of Higgs boson) and $m_t$ (mass of top quark) can also be noticed. At these DM masses, new annihilation channels, $H_0 H_0\rightarrow W^+ W^-$, $H_0 H_0\rightarrow ZZ$, $H_0 H_0\rightarrow h_1 h_1$ and $H_0 H_0\rightarrow t \bar t$ open up. Besides, when $m_{H_0}$ is higher than $m_W$, relic density decreases and becomes underabundant due to the increase in the annihilation cross-section. The DM relic abundance condition is again satisfied at the higher values of DM masses ($m_{H_0}$). For example, in case of BP1 DM particle reaches required relic abundance at $m_{H_0}$=750 GeV.

From the right panels of Figures \ref{fig:2}-\ref{fig:4}, one observes DM direct detection scattering cross-section ($\sigma_{\text{SI}}$) decreases as the DM mass increases. The nature of these plots can be easily understood from Eq. (\ref{eq:37}). The right panels of Figures \ref{fig:2}-\ref{fig:4} show that the calculated $\sigma_{\text{SI}}$ for four BPs are consistent with the bounds given by dark matter direct detection experiments like XENON-1T, LUX, PandaX-II and DarkSide-50. %We checked that for high dark matter masses up to 1 TeV, the dark matter candidate $H_0$ also satisfies the measured relic abundance given by PLANCK. This can be achieved by considering high masses of $m_{h_2}$. 
Thus $H_0$ can be considered as a viable particle candidate of dark matter with the mass of the order of GeV - TeV.


\section{Electroweak Phase transition and Gravitational Waves Production in Extended Inert Doublet Dark Matter Model}
In this section we explore the electroweak phase transition and production of GWs from the considered dark matter model.
\subsection{Effective Potential}
To study the electroweak phase transition (EWPT) in the present model we add the finite temperature correction with the tree-level potential (Eq. (\ref{eq:1})). The finite temperature effective potential can be written as \cite{Wainwright:2012}
\begin{equation}\label{eq:41}
V_{\text{eff}}=V_{\text{}tree-level}+V_{1-loop}^{T=0}+V_{1-loop}^{T\neq0},
\end{equation}
where $V_{1-loop}^{T=0}$ and $V_{1-loop}^{T\neq0}$ are the one-loop corrected potential at zero temperature and at finite temperature respectively. The one-loop effective potential at zero temperature is given by \cite{Wainwright:2012}
\begin{equation}\label{eq:42}
V_{1-loop}^{T=0}=\pm \dfrac{1}{64\pi^2} \sum_{i} n_i m_i^4 \left[ \log\dfrac{m_i^2}{Q^2}-C_i\right],
\end{equation}
where the `+' sign is for bosons and `-' sign is for fermions. Here, $n_i$ is the number of degrees of freedom, $m_i$ is the field-dependent masses of the particles and $i\equiv(h,H_{0},A_{0},H^{\pm},s,W,Z,t$). The degrees of freedom of these particle species are $n_{W^{\pm}}=6$, $n_Z=3$, $n_t=12$ and $n_{h,H_0,A_0,H^+,H^-,s}=1$. The field-dependent masses of the gauge bosons and scalars in terms of $v$ (VEV of $\phi_H$), $v_\phi$ (VEV of $\phi_I$) and $v_s$ (VEV of $S$) at $T=0$ are given by  
\begin{equation}\label{eq:5b}
m_W^2=\dfrac{1}{4}g^2\left(v^2+v_\phi^2\right)\,\,,
\end{equation}

\begin{equation}\label{eq:5c}
m_Z^2=\dfrac{1}{4}\left(g^2+{g^{\prime}}^2\right)\left(v^2+v_\phi^2\right)\,\,,
\end{equation}

\begin{equation}\label{eq:5a}
\mu_{h}^2=m_1^2+3\lambda_1 v^2+ \dfrac{v_\phi^2}{2}\lambda_L+\rho_1 v_s+\rho_2 v_s^2\,\,,
\end{equation}

\begin{equation}\label{eq:6a}
\mu_s^2=m_s^2+\rho_2 v^2+2\rho_3 v_s+3\rho_4 v_s^2+\rho_2^{\prime}v_\phi^2\,\,,
\end{equation}

\begin{equation}\label{eq:8a}
m_{H^\pm}^2=m_2^2+\lambda_2 v_\phi^2+\dfrac{\lambda_3v^2}{2}+\rho_1^\prime v_s+\rho_2^\prime v_s^2\,\,,
\end{equation}

\begin{equation}\label{eq:9a}
m_{H_0}^2=m_2^2+3\lambda_2 v_\phi^2+\dfrac{v^2}{2}\lambda_L+\rho_1^\prime v_s+\rho_2^\prime v_s^2\,\,,
\end{equation}

\begin{equation}\label{eq:10a}
m_{A_0}^2=m_2^2+\lambda_2 v_\phi^2+\left(\lambda_3+\lambda_4-\lambda_5\right)\dfrac{v^2}{2}+\rho_1^\prime v_s+\rho_2^\prime v_s^2\,\,,
\end{equation}

\begin{equation}\label{eq:8b}
m_{G^\pm}^2=m_1^2+\lambda_1 v^2+\dfrac{\lambda_3 v_\phi^2}{2}+\rho_1 v_s+\rho_2 v_s^2\,\,,
\end{equation}

\begin{equation}\label{eq:5b}
m_{G_0}^2=m_1^2+\lambda_1 v^2+ \left(\lambda_3+\lambda_4-\lambda_5\right)\dfrac{v_\phi^2}{2}+\rho_1 v_s+\rho_2 v_s^2\,\,,
\end{equation}
where $g$ and $g^{\prime}$ denote the $SU(2)_L$ and $U(1)_Y$ gauge couplings of the SM respectively. 
%In Eq.(\ref{eq:2}) $\phi$ is the VEV of the doublet $\phi_I$ which inclusion is possible for the neutral component of $\phi_I$ as spontaneous breaking of $Z_2$ is occurred \cite{Blinov:2015vma}.
%before T=0 potential 1 loop
Note that, for $v_\phi=0$, the Eqs. (\ref{eq:5a}) - (\ref{eq:10a}) reduce to the expressions given in Eqs. (\ref{eq:5}) - (\ref{eq:10}).
We work in the Landau gauge ($\xi=0$), where the Goldstone masses become zero at $T=0$ and also there are no ghost contributions \cite{Basler:2016obg}. However, the gauge invariant methods ($\xi\neq0$) \cite{Patel:2011th} can also be applied.
The quantity $Q$ denotes the renormalisable scale which we take $Q=246.22$ GeV in our calculations. However, the selection of $Q$ is not unique. One can vary $Q$ and check the uncertainty due to that in different thermal parameters \cite{Kainulainen:2019kyp,{Laine:2000rm}}. We varied $Q$ for a benchmark point and presented the sensitivity for different parameters in Sect. 5.3. In Eq. (\ref{eq:42}), $C_i$ represents a numerical constant. For $W, \hspace{1mm} Z$ boson the $C_{W,Z}=5/6$ and for the other particles $C_{h,H_{0},A_{0},H^{\pm},s,t}=3/2$. The one-loop effective potential at finite temperature has the form \cite{Wainwright:2012}
\begin{equation}\label{eq:43}
V_{1-loop}^{T\neq0}=\dfrac{T^2}{2\pi^2} \sum_{i} n_i J_{\pm}\left[\dfrac{m_i^2}{T^2}\right],
\end{equation}
where
\begin{equation}\label{eq:44}
J_{\pm}\left(\dfrac{m_i^2}{T^2}\right)=\pm \int_0^{\infty} dy \hspace{1mm} y^2 \log\left(1\mp e^{-\sqrt{y^2+\dfrac{m_i^2}{T^2}}}\right).
\end{equation}
We include the daisy resummation procedure in $V_{1-loop}^{T\neq0}$ \cite{Arnold} by adding the temperature dependent terms to the boson masses. Now, the corrected thermal masses \cite{Huang:2017rzf}, \cite{Blinov:2015vma}-\cite{Gil:2012ya} are
$\mu_{1}^2(T)=m_1^2+c_1 T^2$, $\mu_{2}^2(T)=m_2^2+c_2 T^2$ and $\mu_{3}^2(T)=m_s^2+c_3 T^2$,
where 
\begin{equation}
c_1=\dfrac{6\lambda_1+2\lambda_3+\lambda_4}{12}+\dfrac{3g^2+{g^{\prime}}^2}{16}+\dfrac{1}{6}\rho_2+\dfrac{y_t^2}{4}\,\,,
\end{equation}

\begin{equation}
c_2=\dfrac{6\lambda_2+2\lambda_3+\lambda_4}{12}+\dfrac{3g^2+{g^{\prime}}^2}{16}+\dfrac{1}{6}\rho_2^{\prime}\,\,,
\end{equation}

\begin{equation}
c_3=\dfrac{1}{8}\rho_4+\dfrac{1}{6}\rho_2+\dfrac{1}{6}\rho_2^{\prime}\,\,,
\end{equation}
here $y_t$ denotes the top Yukawa coupling.
In this work we use the CosmoTransitions package \cite{Wainwright:2012} to compute the finite temperature corrections to the tree-level potential.

\subsection{Gravitational Waves Production from Dark Matter}

The bubble nucleation of a true vacuum state (from several metastable states) with surface tension and internal pressure at a temperature known as the nucleation temperature is central to the first-order cosmological phase transition. They can produce different sizes small and large the former types of which tend to collapse whereas the latter types tend to expand after attaining the criticality. These are the latter types, the critical bubbles collide with each other and upon losing their spherical symmetries in the process drive the phase transition and eventual emission of gravitational waves. 

The bubble nucleation rate per unit volume at a particular temperature can be written as \cite{Linde:1983}
\begin{equation}\label{eq:45}
\Gamma=\Gamma_0\left(T \right) e^{-S_3\left(T \right)/T}
\end{equation}
where $\Gamma_0\left(T \right) \propto T^4$, $S_3\left(T \right)$ is the Euclidean action of the critical bubble. The Euclidean action $S_3\left(T \right)$ can be expressed as \cite{Linde:1983}
\begin{equation}\label{eq:46}
S_{3}=4\pi \int dr \hspace{1mm}r^{2} \left[ \dfrac{1}{2} \left(\partial_{r} \vec{\phi} \right)^2 +V_{eff}\right],
\end{equation}
where 
%$\vec{\phi}$ represents a vector of scalar fields and 
$V_{eff}$ is the effective finite temperature potential represented in Eq. (\ref{eq:41}). Nucleation of the bubble occurs at the nucleation temperature $T_n$ if it satisfies the condition $S_3\left(T_n \right)/T_n\approx 140$ \cite{Wainwright:2012}.

The gravitational waves are produced from the first-order phase transition mainly by the three mechanisms namely bubble collisions \cite{Kosowsky:1992a}-\cite{Caprini:2008}, sound wave \cite{Hindmarsh:2014}-\cite{Hindmarsh:2015} and turbulence in the plasma \cite{Caprini:2006}-\cite{Caprini:2009}. The total GW intensity $\Omega_{\text{GW}}{\rm h}^2$ as a function of frequency can be expressed as the sum of the contributions from the three components \cite{Kosowsky:1992a}-\cite{Caprini:2009}
\begin{equation}\label{eq:47}
\Omega_{\text{GW}}{\rm h}^2=\Omega_{\text{col}}{\rm h}^2+ \Omega_{\text{SW}}{\rm h}^2+ \Omega_{\text{turb}}{\rm h}^2.
\end{equation}
The component from the bubbles collision $\Omega_{\text{col}}{\rm h}^2$ is given by
\begin{equation}\label{eq:48}
\Omega_{\text{col}}{\rm h}^2=1.67\times 10^{-5} \left(\dfrac{\beta}{H} \right) ^{-2} \dfrac{0.11 v_{w}^3}{0.42+v_{w}^2} \left(\dfrac{\kappa \alpha}{1+\alpha}\right)^2 \left(\dfrac{g_*}{100}\right)^{-\frac{1}{3}}\dfrac{3.8 \left(\dfrac{f}{f_{col}}\right)^{2.8}}{1+2.8 \left(\dfrac{f}{f_{\text{col}}}\right)^{3.8}}\,\,\,,
\end{equation}
with the parameter $\beta$
\begin{equation}\label{eq:49}
\beta=\left[H T \dfrac{d}{dT}\left( \dfrac{S_3}{T}\right) \right]\bigg|_{T_n},
\end{equation}
where $T_n$ is the nucleation temperature and $H_n$ is the Hubble parameter at $T_n$. A conservative estimate of the bubble wall velocity $v_w$ can be expressed as \cite{Kamionkowski:1994, {Chao:2017vrq},{Dev:2019njv},{Steinhardt:1982}}
\begin{equation}\label{eq:50}
v_w=\dfrac{1/\sqrt{3}+\sqrt{\alpha^2+2\alpha/3}}{1+\alpha}.
\end{equation}
Here in this work, we use the above mentioned expression (Eq. (\ref{eq:50})) for calculating the bubble wall velocity. However, in some literatures, $v_w$ is taken to be 1 for simplicity (\cite{Shajiee:2018jdq, {Mohamadnejad:2019vzg}, {Chala:2019rfk}}). A more detailed discussion on the choice of $v_w$ can be found in \cite{Kozaczuk:2015owa} which results in $v_w$ $\gtrsim$ 0.2. The parameter $\kappa$ in Eq. (\ref{eq:48}) represents the fraction of latent heat deposited in a thin shell which is given by
\begin{equation}\label{eq:51}
\kappa=1-\dfrac{\alpha_{\infty}}{\alpha},
\end{equation}
with \cite{Shajiee:2018jdq,{Caprini:2015zlo}}
\begin{equation}\label{eq:52}
\alpha_{\infty}=\dfrac{30}{24\pi^2 g_{*}} \left(\dfrac{v_n}{T_n} \right)^2 \left[6 \left( \dfrac{m_W}{v}\right)^2 +3\left( \dfrac{m_Z}{v}\right)^2 +6\left( \dfrac{m_{t}}{v}\right)^2\right].
\end{equation}
In the above, $v_n$ is the vacuum expectation value of Higgs at $T_n$ and $m_W$, $m_Z$ and $m_t$ are the masses of W, Z and top quarks respectively. The parameter $\alpha$ is defined as the ratio of vacuum energy density $\rho_{\text{vac}}$ released by the electroweak phase transition to the background energy density of the plasma $\rho_*^{\text{rad}}$ at $T_n$. The expression of $\alpha$ has the form
\begin{equation}\label{eq:53}
\alpha=\left[\dfrac{\rho_{\text{vac}}}{\rho^*_{\text{rad}}}\right]\bigg|_{T_n}.
\end{equation}
with
\begin{equation}\label{eq:54}
\rho_{\text{vac}}=\left[\left(V_{\text{eff}}^{\text{high}}-T\dfrac{dV_{\text{eff}}^{\text{high}}}{dT} \right)-\left(V_{\text{eff}}^{\text{low}}-T\dfrac{dV_{\text{eff}}^{\text{low}}}{dT} \right)\right],
\end{equation}
and
\begin{equation}\label{eq:55}
\rho^*_{\text{rad}}=\dfrac{g_* \pi^2 T_n^4}{30}.
\end{equation}
The parameter $\rho_{vac}$ can also be calculated using the trace anomaly \cite{Hindmarsh:2015}, where the same requires an additional factor of $1/4$ in front of the term $T\dfrac{dV}{dT}$ (Eq. (\ref{eq:54})).
The quantity $f_\text{col}$ in Eq. (\ref{eq:48}) is the peak frequency produced by the bubble collisions which takes the form
\begin{equation}\label{eq:56}
f_{\text{col}}=16.5\times10^{-6}\hspace{1mm} \text{Hz} \left( \dfrac{0.62}{v_{w}^2-0.1 v_w+1.8}\right)\left(\dfrac{\beta}{H} \right) \left(\dfrac{T_n}{100 \hspace{1mm} \text{GeV}} \right) \left(\dfrac{g_*}{100}\right)^{\frac{1}{6}}.
\end{equation}
The sound wave (SW) component of the gravitational wave (Eq. (\ref{eq:47})) is given by
\begin{equation}\label{eq:57}
\Omega_{\text{SW}}{\rm h}^2=2.65\times 10^{-6} \left(\dfrac{\beta}{H} \right) ^{-1} v_{w} \left(\dfrac{\kappa_{v} \alpha}{1+\alpha}\right)^2 \left(\dfrac{g_*}{100}\right)^{-\frac{1}{3}}\left(\dfrac{f}{f_{\text{SW}}}\right)^{3} \left[\dfrac{7}{4+3 \left(\dfrac{f}{f_{\text{SW}}}\right)^{2}}\right]^{\frac{7}{2}},
\end{equation}
where $\kappa_v$ is the faction of latent heat transformed into the bulk motion of the fluid which has the following form
\begin{equation}\label{eq:58}
\kappa_v=\dfrac{\alpha_{\infty}}{\alpha}\left[ \dfrac{\alpha_{\infty}}{0.73+0.083\sqrt{\alpha_{\infty}}+\alpha_{\infty}}\right].
\end{equation}
In Eq. (\ref{eq:57}) $f_{\text{SW}}$ denotes the peak frequency produced by the sound wave mechanisms which takes the form

\begin{equation}\label{eq:59}
f_{\text{SW}}=1.9\times10^{-5}\hspace{1mm} \text{Hz} \left( \dfrac{1}{v_{w}}\right)\left(\dfrac{\beta}{H} \right) \left(\dfrac{T_n}{100 \hspace{1mm} \text{GeV}} \right) \left(\dfrac{g_*}{100}\right)^{\frac{1}{6}}.
\end{equation}
The component from the turbulence in the plasma $\Omega_{\text{turb}}{\rm h}^2$ is given by
\begin{equation}\label{eq:60}
\Omega_{\text{turb}}{\rm h}^2=3.35\times 10^{-4} \left(\dfrac{\beta}{H} \right) ^{-1} v_{w} \left(\dfrac{\epsilon \kappa_v \alpha}{1+\alpha}\right)^{\frac{3}{2}} \left(\dfrac{g_*}{100}\right)^{-\frac{1}{3}} \dfrac{\left(\dfrac{f}{f_{\text{turb}}}\right)^{3}\left( 1+\dfrac{f}{f_{\text{turb}}}\right)^{-\frac{11}{3}}}{\left(1+\dfrac{8\pi f}{h_{*}}\right)},
\end{equation}
where $\epsilon=0.1$ and $f_{\text{turb}}$ denotes the peak frequency produced by the turbulence mechanism which can be written as
\begin{equation}\label{eq:61}
f_{\text{turb}}=2.7\times10^{-5}\hspace{1mm} \text{Hz} \left( \dfrac{1}{v_{w}}\right)\left(\dfrac{\beta}{H} \right) \left(\dfrac{T_n}{100 \hspace{1mm} \text{GeV}} \right) \left(\dfrac{g_*}{100}\right)^{\frac{1}{6}}.
\end{equation}
In Eq. (\ref{eq:60}) the parameter $h_*$ has the following form
\begin{equation}\label{eq:62}
h_{*}=16.5\times10^{-6}\hspace{1mm} \text{Hz} \left(\dfrac{T_n}{100 \hspace{1mm} \text{GeV}} \right) \left(\dfrac{g_*}{100}\right)^{\frac{1}{6}}.
\end{equation}

We repeat our analysis by considering an updated expression of $\kappa_v$ \cite{Ellis:2018mja,{Espinosa:2010hh}}, where the prefactor $\dfrac{\alpha_{\infty}}{\alpha}$ is taken to be 1 and furnish our results in Sect. 5.3. The updated expression of $\kappa_v$ is given by \cite{Ellis:2018mja,{Espinosa:2010hh}}

\begin{equation}\label{eq:62a}
\kappa_v\simeq\left[ \dfrac{\alpha_{\infty}}{0.73+0.083\sqrt{\alpha_{\infty}}+\alpha_{\infty}}\right].
\end{equation}
In this work, Eqs. (\ref{eq:47})-(\ref{eq:62a}) are used for the computation of gravitational wave intensity.

\subsection{Calculations and Results}
As mentioned earlier the computation of GW intensity from the first-order phase transition in the present particle DM model (inert doublet and a scalar singlet extended SM) has been performed for four chosen benchmark points for the model parameters given in Table 1. 

We calculate gravitational wave intensity from the model and compare it with the sensitivity curves of different GW detectors such as BBO, eLISA, ALIA, DECIGO, U-DECIGO and aLIGO. The GW intensity depends mainly on factors like strength of the first-order phase transition (the parameter $\alpha$), the time-scale of the phase transition (the parameter $1/\beta$), bubble wall velocity $v_{w}$, nucleation temperature $T_n$ and Higgs VEV $v_n$ at the nucleation temperature $T_n$. In order to calculate the GW intensity, we first calculate the transition temperature of the first-order phase transition. In calculating this, the finite temperature effective potential (Eq. (\ref{eq:41})) is first computed. For these calculations we use a publicly available package namely the Cosmotransition package \cite{Wainwright:2012}. The tree-level potential (Eq. (\ref{eq:1})) serves as an input to this package and provides the parameters related to the phase transition. The GW intensity is estimated by using Eqs. (\ref{eq:47})-(\ref{eq:62a}). In this work, we have selected four BPs (Table \ref{t1}) such that they satisfy all constraints mentioned in Sect. 3. The relic abundance $\Omega_{\text{DM}}\text{h}^2$ and the direct detection scattering cross-section $\sigma_{\text{SI}}$ obtained from each of the four BPs are also given in Table \ref{t1}.

\begin{figure}[H]
\centering
\includegraphics[width=8cm,height=5.5cm]{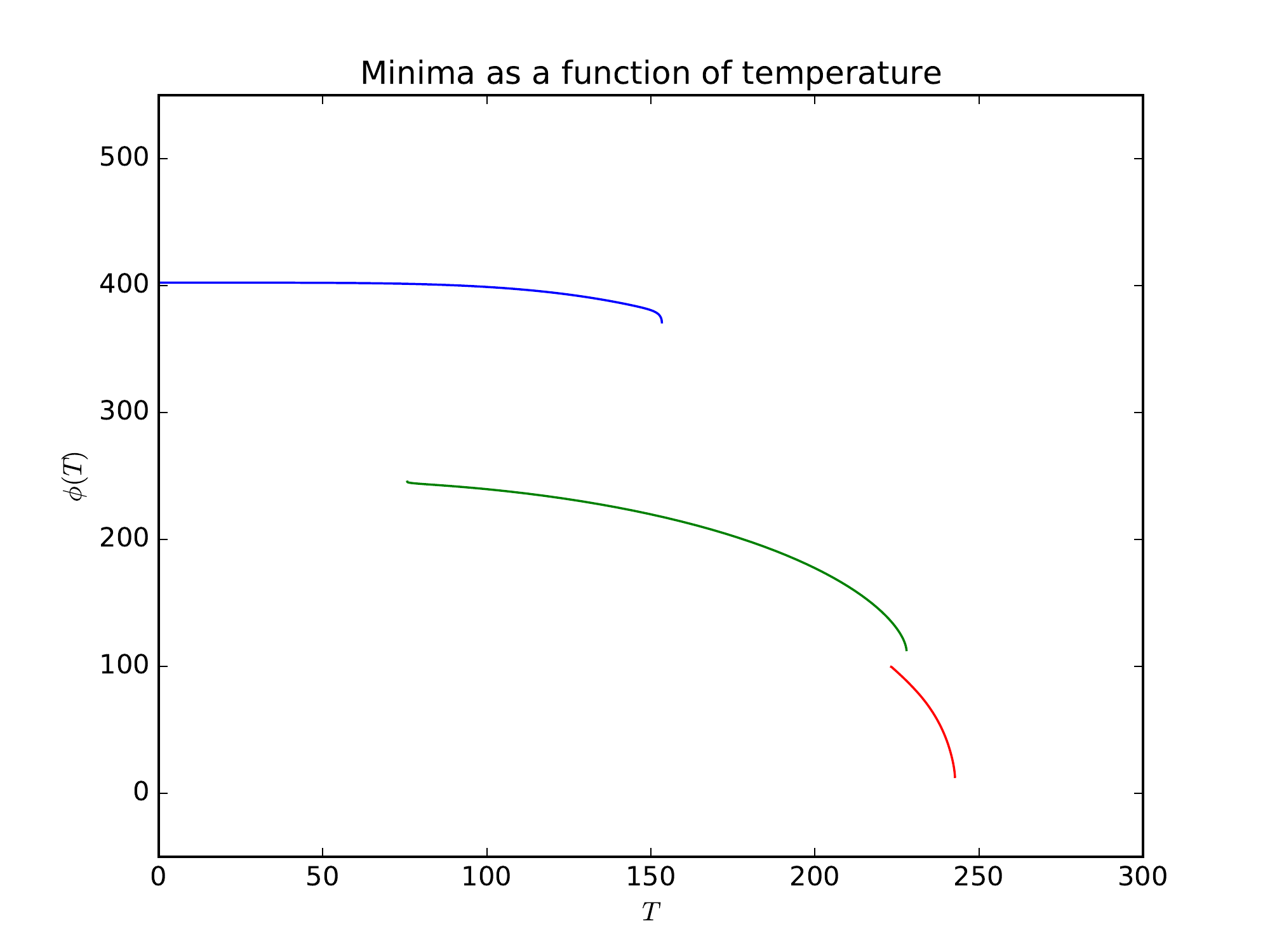}
\includegraphics[width=8cm,height=5.5cm]{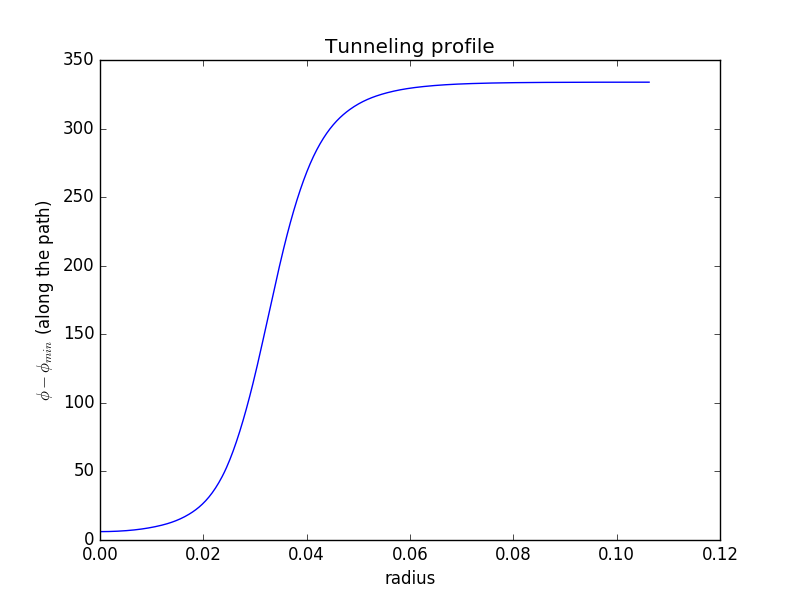}
\caption{Phase transition properties for the BP1. The left panel is for the position of the minima as a function of temperature and the right panel is for the tunneling profile as a function of bubble radius.}
\label{fig:5}
\end{figure}

In Figure \ref{fig:5} we show the phase transition properties for BP1. The left and right panels of Figure \ref{fig:5} show the phase structure of the model and the tunneling profile as a function of the bubble radius respectively. As can be seen in the left panel of Figure \ref{fig:5}, there exist two transition temperatures at $T_n=119.86$ GeV and $T_n=226.55$ GeV. We only consider the lower of the two as the phase transition temperature because the low nucleation temperature is more sensitive to probe the GW signal. An electroweak phase transition occurs when the temperature of the Universe drops which results in a separation of potential between a high phase and a low phase by a potential barrier. In this case, a first-order phase transition occurs at the nucleation temperature $T_n=119.86$ GeV. The phase transition properties for BP2, BP3 and BP4 are also studied but no significant differences from BP1 are noted. The computed values of the parameters ($v_n, \hspace{1mm}T_c,\hspace{1mm}T_n,\hspace{1mm}\alpha,\hspace{1mm}\beta/H$) corresponding to each of the four BPs to be uesd for the calculations of GW intensity are furnished in Table \ref{t2}. 
\begin{table}[H]
\centering
\small
\begin{tabular}{|l|c|c|c|c|c|c|c|c|r|}
\hline
BP&$v_n$&$T_c$&$T_{n}$&$\alpha$&$\dfrac{\beta}{H}$\\
 &in GeV&in GeV&in GeV&&\\
 \hline
1&226.89&135.68&119.86&0.24&317.86\\
\hline
2&191.03&146.89&132.14&0.25&402.89\\
\hline
3&14.22&130.78&116.99&0.23&692.01\\
\hline
4&209.95&170.92&158.24&0.19&783.65\\
\hline
\end{tabular}
\caption{The values for the parameters used to calculate the GW intensity for each of the chosen BPs. See text for details.}\label{t2}
\end{table}
From Table \ref{t2}, one can see that the nucleation temperature $T_n$ is smaller than the critical temperature $T_c$ (the temperature at which there exist two degenerate minima separated by a potential barrier) for each of the BPs. Since, the selection of $Q$ is not unique, we have varied $Q$ form 68 GeV to 270 GeV and found that the parameters have the following uncertainty: 17.6 \%, 14.8 \% and 0.9 \% for $T_c$,  $T_n$ and $v_n$ respectively. This indicates that the thermal parameters can vary significantly with $Q$.

\begin{figure}[H]
\centering
\includegraphics[width=8.55cm,height=8cm]{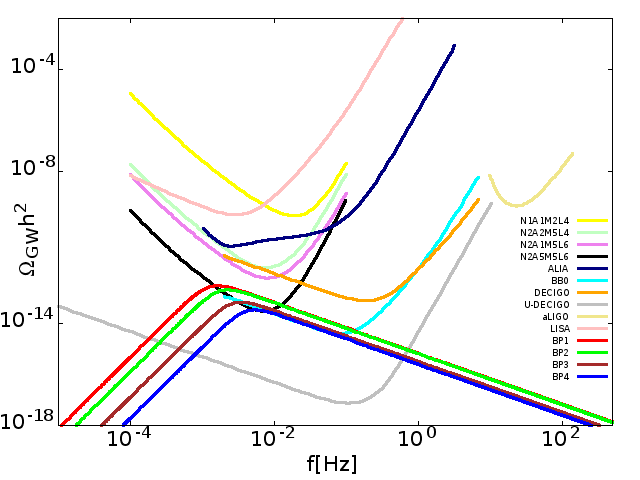}
\includegraphics[width=8.55cm,height=8cm]{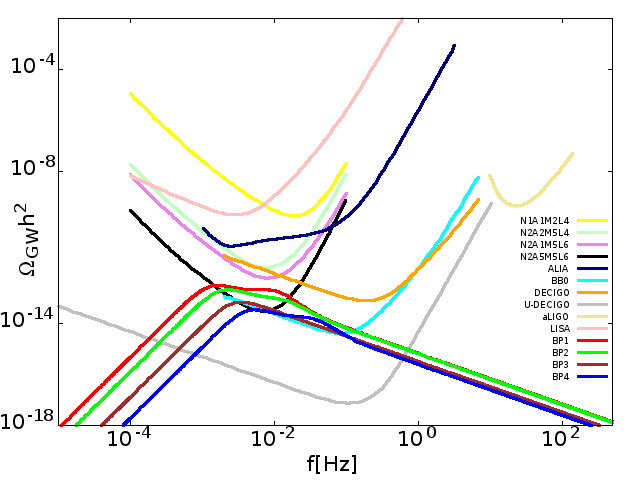}
 \caption{Comparison of the chosen four BPs with the sensitivity curves of N1A1M2L4, N2A2M5L4, N2A1M5L6 and N2A5M5L6 configurations of eLISA, ALIA, BBO, DECIGO, U-DECIGO, aLIGO and LISA detectors. The left and right panels show the change in the GW intensity vs frequency using the expression for $\kappa_v$ following Eq. (\ref{eq:58}) and Eq. (\ref{eq:62a}) respectively.}
\label{fig:6}
\end{figure}

Using the computed numerical values for each of the four sets given in Table \ref{t2} (corresponding to the BPs given in Table \ref{t1}) the GW intensities and frequencies from first-order phase transitions are now calculated from Eqs. (\ref{eq:47})-(\ref{eq:62a}) for each of the sets. In order to test whether the sound wave contribution to the overall GW signal is significant, the suppression factor ($H R_*/\bar{U}_f$, where $R_*$ is the mean bubble separation and $\bar{U}_f$ is the root-mean-
square (RMS) fluid velocity) is calculated \cite{Caprini:2015zlo,{Ellis:2018mja},{Ellis:2019oqb}}. In our model, the sound wave contribution of the overall GW amplitude turns out to be 1.76 ($>$ 1) for BP1. This implies that the sound wave contribution, in this case, survives more than a Hubble time (long-lasting sound waves) and hence contribute significantly. The applicability of this issue can be tested through a dedicated numerical lattice simulation. The variations of GW intensities with the frequencies for different BPs are shown in left and right panels of Figure \ref{fig:6}, where the GW intensities for different BPs are calculated using the expressions of $\kappa_v$ following Eq. \ref{eq:58} and \ref{eq:62a} respectively. The sensitivity curves of GW detectors (BBO, eLISA, ALIA, DECIGO, U-DECIGO, aLIGO and LISA) are also shown in both of the panels of Figure \ref{fig:6} for comparison. One of the ways to graphically represent the sensitivity of a GW detector is called the power-law-integrated sensitivity curve \cite{Thrane:2013oya} which calculates power across all frequencies in the sensitivity band by integrating the noise-weighted signal over frequency. However, we have presented our results in such a way that the sensitivity curves for different GW detectors can directly be compared with the stochastic backgrounds and contribution from other types of sources by following Ref. \cite{Moore:2015}. The future generation space-based GW detectors such as eLISA, BBO, ALIA, and DECIGO are expected to explore the frequency range from millihertz to decihertz. These space missions complement the effort of future generation ground-based GW detector aLIGO by providing the low frequency coverage with much higher sensitivity. This is because of the longer arms of the space-based detectors and absence of the seismic noise. The combined effort of the future space and ground-based telescopes, operational at different energies (eLISA with four different configurations such as N1A1M2L4, N2A2M5L4, N2A1M5L6 and N2A5M5L6 - millihertz band, ALIA, BBO, DECIGO - decihertz band, and aLIGO - decahertz band) will provide unprecedented detection of GWs. In addition, the recently proposed variant of DECIGO based on Fabry-Perot interferometry, is expected to significantly reduce the signal-to-noise ratio of GW backgrounds. As a result of which the best sensitivity in the frequency range 0.1 hertz to 10 hertz will be around $\Omega_{\text{GW}}\text{h}^2 = 10^{-16} - 10^{-15}$. It can be seen from Figure \ref{fig:6} that for different BPs, GW intensities attain peaks at different frequencies. The peak of the GW intensity for BP1, BP2, BP3, and BP4 appears at frequencies $1.7\times 10^{-3}$ Hz, $2.3 \times 10^{-3}$ Hz, $3.5 \times 10^{-3}$ Hz and $5.4\times 10^{-3}$ Hz respectively. We obtained higher GW intensity for BP1 as compared to the other BPs. As seen from Figure \ref{fig:6} the GW intensities for all the BPs (BP1, BP2, BP3 and BP4) lie within the sensitivity curves of the N2A5M5L6 configuration of eLISA, BBO and U-DECIGO. 
%whereas the results corresponding to BP4 lie within the sensitivity curves of the BBO and U-DECIGO.
However, BP1 is special because it shows higher intensity than the rest of the cases. It may be inferred from Figure \ref{fig:6} and Table \ref{t2} that the GW intensity mainly depends on $\beta$. For BP1, the value of the parameter $\beta$ is the smallest and the corresponding GW intensity is the highest along with the lowest peak frequency at $1.7\times 10^{-3}$. The choice of Eq. (\ref{eq:62a}) over Eq. (\ref{eq:58}) introduces slight increase in the GW intensity shown in the right panel of Figure \ref{fig:6} as compared to the left panel. Small increase in the falling part of the GW spectrum is observed for three BPs namely BP1, BP2 and BP4 are attributed to the marginal increase of the GW intensity due to the sound wave contribution. However, the primary peak intensities for different BPs remain the same for both of the panels.

\section{Summary and Conclusions}
In this work, we have explored the possible production of GWs from the first-order phase transition of the early Universe from a viable particle dark matter model and its detectability with the future GW detectors (BBO, eLISA, ALIA, DECIGO, U-DECIGO and aLIGO). We discussed the first-order phase transition that may be initiated by the present extended SM where the scalar sector of the SM is modified by adding an additional inert scalar doublet and a real scalar singlet. A discrete $Z_2$ symmetry has been imposed on the scalar potential which enables the lightest stable inert particle to be a viable particle candidate of dark matter. Due to the imposition of the $Z_2$ symmetry, the inert particle does not interact with the SM particles and also it does not acquire any VEV. The added real singlet scalar acquires a VEV on spontaneous symmetry breaking and it mixes with the SM Higgs which resulted in two new physical scalar eigenstates $h_1$ and $h_2$. Here we considered $h_1$ as the SM like Higgs boson and the other scalar $h_2$ as the added physical scalar. We constrain the model parameters using the conditions for vacuum stability, perturbativity, LEP and LHC bounds, dark matter relic density as measured by PLANCK and the bounds from the direct detection experiments. The future collider sensitivity is also used to constrain the scalar mixing angle and check for viability of the model with DM phenomenology and Gravitational Wave experiments. From the model parameter space thus constrained, we choose four benchmark points from the constrained parameter space for the computation of GW intensity and explore the GW production from the first-order phase transition of the tree-level potential. We include the finite temperature corrections of the tree-level potential. We calculate the GW intensity for four BPs. It has been found that the GW intensity increases as $\beta$ (Eq. (\ref{eq:49})) decreases. In addition, the lower value of $\beta$ also lowers the frequency at which the maximum GW intensity is produced. We then explore whether the GW results for the four BPs lie within the sensitivity range of the GW detectors and found that the GW signals are detectable by the following future generation detectors namely BBO, U-DECIGO and eLISA (configuration - N2A5M5L6) within the detectable range. In this work, we show that the extended inert doublet model under consideration can explain the dark matter as well as the strong first-order electroweak phase transition. Thus we found that our model constrained by limits from future collider experiments, is consistent with DM searches and also in agreement with the sensitivity of GW detectors. The results also imply that the detection of GW signals by future spaceborne detectors could also be useful to ascertain the nature of particle dark matter candidates.

\vspace{5mm}
\noindent {\bf Acknowledgments}

The authors would like to thank A. Dutta Banik and A. Biswas for useful discussions.

{}
\end{document}